\documentclass[doublecol]{epl2}
\usepackage{amsmath}
\usepackage{amssymb}
\usepackage{graphicx}

\title{Baryon asymmetry resulting from a quantum phase
transition in the early universe} \shorttitle{Baryon asymmetry}
\author{V. R. Shaginyan \inst{1,2}\thanks {Email:
\email{vrshag@thd.pnpi.spb.ru}} \and G. S. Japaridze\inst{2} \and
M. Ya. Amusia\inst{3} \and A. Z. Msezane\inst{2} \and K. G.
Popov\inst{4}} \shortauthor{V. R. Shaginyan \etal}
\institute{\inst{1} Petersburg Nuclear
Physics Institute, Gatchina, 188300, Russia\\
\inst{2} Clark Atlanta University, Atlanta, GA 30314, USA\\
\inst{3} Racah Institute of Physics, Hebrew University, Jerusalem
91904, Israel\\ \inst{4} Komi Science Center, Ural Division, RAS,
Syktyvkar,167982, Russia}


\abstract{A novel mechanism for explaining the matter-antimatter
asymmetry of the universe is considered. We assume that the
universe starts from completely symmetric state and then, as it
cools down, it undergoes a quantum-phase transition which in turn
causes an asymmetry between matter and anti-matter. The mechanism
does not require the baryon number violating interactions or $CP$
violation at a microscopic level. Our analysis of the
matter-antimatter asymmetry is in the context of conspicuous
experimental results obtained in the condensed-matter physics.}

\begin{document}
\maketitle

One of the important and long standing problems of modern cosmology
and astrophysics is the matter-antimatter asymmetry: the observable
part of the universe contains mostly baryons and antibaryons
produced locally as the byproduct of nuclear reactions
\cite{rev1,rev2,rev3,rev4}. For the globally symmetric universe one
can put a strong constraint on the size $l$ of antimatter clusters
\cite{rujula}, $l\,>\,1000$ Mpc. This number may be compared with
the visible size of the universe, 3000 Mpc.

A convenient dimensionless number which characterizes the magnitude
of the baryon asymmetry is the ratio of the baryonic charge density
$(n_{B}-n_{\bar{B}})$ to the density of the cosmic microwave
background $n_{\gamma}$ \cite{BBB}
\begin{equation}
\label{BB} {n_{B}-n_{\bar{B}}\over n_{\gamma}}\leq 3\cdot 10^{-10}
\end{equation}
Estimate (\ref{BB}), excluding antimatter on scales of order $\sim 20$
Mpc \cite{BBB}, may be obtained assuming that at earlier times, at
temperatures well above $100$ MeV, the universe had one extra quark
per about $10^{10}$ quark-antiquark pairs and this tiny excess is
responsible for the entire baryonic matter in the present universe.

Although there is no logical contradiction to assume that an excess
of quarks over antiquarks is built in as an initial condition, this
{\it ad hoc} hypothesis can not be justified within the
inflationary scenario which does not provide such an initial
condition \cite{rev1,rev2,rev3,rev4}. Thus, it becomes necessary to
explain baryon asymmetry without introducing the minute particle
excess at the initial stage of the big bang.

Sakharov \cite{sakh} was first to formulate the conditions
necessary for generating the observed baryon asymmetry from an
initially symmetric state. These are: the baryon number
non-conservation, $C$ and $CP$ violation, and a departure from the
thermal equilibrium. Numerous possible scenarios incorporating
Sakharov's conditions readily followed. The detailed discussion of
these mechanisms such as electroweak baryogenesis, baryonic charge
condensate, baryogenesis via lepto genesis, baryogenesis through
evaporation of primordial black holes, out of equilibrium decays of
massive particles, baryon number non conservation caused by the
triangle anomaly in baryonic current and baryogenesis in the
presence of spontaneously broken Lorentz symmetry can be found in
\cite{rev1,rev2,rev3,rev4}. It is worth noting that all these
scenarios require extension of the standard model of particle
physics.

In this letter we propose a mechanism for explaining the
matter-antimatter asymmetry of the universe, which does not require
any extension of the standard model of particle physics or the
standard model of cosmology. Our approach is based on the
observation that the condensed-matter physics of strongly
correlated Fermi systems and topologically protected gapless
fermions without dispersion forming the flat bands may offer
opening for designing such a mechanism \cite{vol,vol1,vol2}. We
propose that the universe began from a completely symmetric state
with the baryon number and $CP$ conserved at the end of inflation
when the particle production started. The observed asymmetry may be
explained by suggesting that in the post inflation epoch, at
baryogenesis, as the universe cooled down in approximately 10
orders of magnitude, it came near a quantum phase transition. At
that time an excess of matter over the antimatter in the universe
was generated. This phase transition could wash out the antibaryons
from the universe ground state wave function. Such a quantum phase
transition can be represented by the fermion condensation quantum
phase transition (FCQPT) that does not support quasiparticle-hole
symmetry \cite{ks,ksk,pr,vrshag}. We note that flat bands related
to FCQPT were observed in 2+1 dimensional quantum field theory
which is dual to a gravitational theory in the anti-de Sitter
background \cite{lee}. For the detailed discussion of novel
features exhibited by the strongly correlated Fermi systems see
review \cite{pr}.

Our suggestion is based on the results of theoretical
\cite{pr,vrshag,obz} and experimental \cite{deuts,exp1,exp2}
studies of novel systems of condensed matter physics - the strongly
correlated Fermi systems and the quantum phase transitions within.
These systems do not belong to the well known class of Landau Fermi
Liquid \cite{landau}, and exhibit the non-Fermi Liquid behavior
strikingly different from those of the Landau Fermi-Liquid
\cite{ste}.

We propose that the universe exhibits non-Fermi Liquid behavior and
therefore shares the features of non-Fermi Liquid, in particular
the spontaneous breakdown of quasiparticle-hole symmetry. The
quasiparticle-hole asymmetry manifests itself on a macroscopic
scale as an asymmetric conductivity \cite{deuts,exp1,exp2}. This
phenomenon of quasiparticle-hole asymmetry serves as  the guiding
principle in our suggestion of explaining matter-antimatter
asymmetry from the results of condensed matter physics of strongly
correlated fermi systems. Below we briefly discuss the
quasiparticle and hole properties in Landau Fermi-Liquid and
non-Fermi Liquid which are necessary to illustrate our suggestion
and provide the basis for interpreting the observed baryon
asymmetry as resulting from a quantum phase transition. Throughout
holes will represent the matter (baryons) and quasiparticles will
serve as an analogy of antimatter (antibaryons).

As it is well known, the basic thermodynamic and transport
properties of Landau Fermi-Liquid are described in terms of
quasiparticles - the weakly excited states over the Fermi sea (or
the Fermi level $E_F$) \cite{landau}. Landau Fermi Liquid is
symmetric with regard to quasiparticles and holes. The latter are
the ``mirror images'' of quasiparticles with the same mass but
opposite charge; in particle physics terminology the quasiparticles
above and the holes below the Fermi sea are fermions above and
antifermions below the Dirac sea. Microscopic Hamiltonian
describing Landau Fermi-Liquid is fully symmetric with regards to
holes and quasiparticles, and this ``matter-antimatter symmetry''
holds on a macroscopic scale as well \cite{landau}.

\begin{figure}[!ht]
\begin{center}
\includegraphics[width=0.47\textwidth]{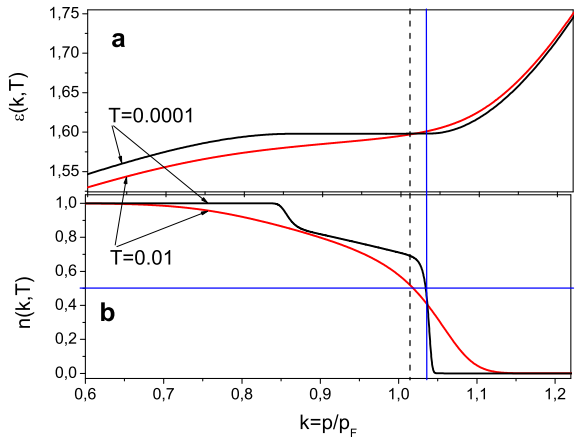}
\end{center}
\caption{The single particle energy $\varepsilon({\bf k},T)$ ({\bf
a}) and the distribution function $n({\bf k},T)$ ({\bf b}) at
finite temperatures as functions of the dimensionless variable
$k=p/p_F$. Temperature is measured in units of $E_F$. At $T=0.01$
and $T=0.0001$ the vertical dashed and solid lines respectively
show the position of the Fermi level $E_F$ at which $n({\bf
k},T)=0.5$ as depicted by the horizontal line. As $T\to0$ and
consistent with eq. \eqref{GS}, the single particle energy
$\varepsilon({\bf k},T)$ becomes more flat in the region
$(p_f-p_i)$ and the distribution function $n({\bf k},T)$ in this
region becomes more asymmetric with respect to the Fermi level
$E_F$ producing the quasiparticle-hole asymmetry related to the
non-Fermi Liquid behavior.} \label{figfc}
\end{figure}

The theory of Landau Fermi Liquid is based on a representation of the system as a
gas of interacting quasiparticles, the number of which is equal to
the number $N$ of particles \cite{landau}. The ground state energy
$E$ of a uniform Fermi system is treated as a functional of the
quasiparticle distribution $n({\bf p})$. Under arbitrary variation
of $n({\bf p})$, conserving the particle number density $x$, the energy
$E$ is changed according to the formula \cite{ks,ksk}
\begin{equation}\label{GS}
\delta E=\int(\varepsilon({\bf p})-\mu)\delta n({\bf p})\frac{d{\bf
p}}{(2\pi)^3}.
\end{equation}
Here $\varepsilon({\bf p})=\delta E/\delta n({\bf p})$ is the
energy of a quasiparticle and $\mu$ is the chemical potential.
Distribution $n({\bf p})$ at $T=0$ is a Fermi step $n(p)=\theta(p-p_F)$,
$p_F$ is the Fermi momentum, and $x=p_F^3/3\pi^2$. At finite
temperatures $n({\bf p})$ is given by the Fermi-Dirac distribution
\begin{equation} n({\bf p},T)=
\{1+\exp[{(\varepsilon({\bf p},T)-\mu)}/ {T}]\}^{-1}.\label{n}
\end{equation}
A necessary stability condition for Landau Fermi-Liquid is that the
group velocity of quasiparticles be nonnegative at the Fermi
surface: $v_g(p)=d\varepsilon(p)/dp\geq0$. In that case $\delta
E>0$ and we obtain from eq. \eqref{GS} that $\varepsilon(p)=\mu$.
This condition can be reformulated as an equation of the minimum
\cite{ks,ksk}
\begin{equation} \frac{\delta E}{\delta
n({\bf p})}=\varepsilon({\bf p})=\mu; \, p_i\leq p\leq p_f.
\label{FL8}\end{equation} The distribution function found from eq.
\eqref{FL8} differs from the step function in the interval from
$p_i$ to $p_f$; inside this interval $0<n({\bf p})<1$. Outside this
interval $n({\bf p})$ coincides with the step function. When the
density $x\to x_{FC}$ where $x_{FC}$ is a quantum critical point
(QCP) of FCQPT, a nontrivial solution of eq. \eqref{FL8} emerges.
At finite $T$, a solution $n({\bf p})$ of eq. (\ref{FL8}) and the
corresponding single-particle spectrum $\varepsilon({\bf p})$ are
depicted in fig. \ref{figfc} \cite{obz}.
At the Fermi level $\varepsilon({\bf p},T)=\mu$, then from eq.
\eqref{n} the distribution function $n({\bf p},T)=1/2$. The
vertical dashed and solid lines in fig. \ref{figfc} crossing the
distribution function at the Fermi level illustrate the asymmetry
of the corresponding distribution functions with respect to the
Fermi level at $T=0.01$ and $T=0.0001$ respectively. As $T\to0$ as
seen from fig. \ref{figfc}, the number density of holes
$H=\sum_{\varepsilon(p)<E_F}(1-n(p))$ is finite, while the number
density of quasiparticles $P=\sum_{\varepsilon(p)>E_F}n(p)$
vanishes. Clearly the solutions of eq. \eqref{FL8} strongly violate
the particle-hole symmetry, and the asymmetry $R_A=(H-P)/(H+P)$
increases becoming more pronounced at diminishing temperatures.

\begin{figure} [! ht]
\begin{center}
\includegraphics [width=0.47\textwidth]{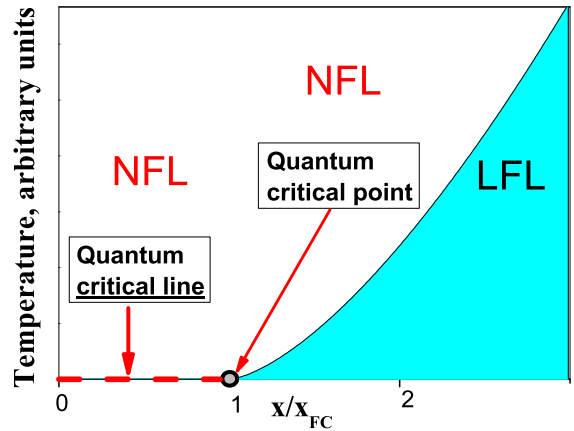}
\end{center}
\vspace*{-0.8cm} \caption{Schematic phase diagram of a system with
FCQPT. The number density $x$ is taken as the control parameter and
depicted as $x/x_{FC}$. The quantum critical point at $x/x_{FC}=1$
of FCQPT is denoted by the arrow. At $x/x_{FC}>1$ and sufficiently
low temperatures, the system is in the Landau Fermi-Liquid state as
shown by the shadow area. At finite temperatures and beyond the
critical point, $x/x_{FC}<1$, the system is above the quantum
critical line depicted by the dashed line and shown by the vertical
arrow. The location of the system in the NFL region is
characterized by the quasiparticle-hole asymmetry related to the
non-Fermi Liquid behavior.}\label{fig1}
\end{figure}
A schematic phase diagram of the system which is driven to FCQPT by
variation of $x$ is reported in fig. \ref{fig1}. Upon approaching
QCP of FCQPT at $x=x_{FC}$, the system remains in the Landau Fermi
Liquid region at sufficiently low temperatures as is shown by the
shadow area. The temperature range of the shadow area shrinks as
the system approaches QCP. At $x_{FC}$ shown by the arrow in fig.
\ref{fig1}, the system demonstrates the non-Fermi Liquid behavior
down to the lowest temperatures. Below QCP at finite temperatures
the behavior remains the non-Fermi Liquid one with the
particle-hole asymmetry. In that case as $T\to 0$, the system is
approaching a quantum critical line (shown by the vertical arrow
and the dashed line in fig. \ref{fig1}) rather than a quantum
critical point. It is seen from fig. \ref{fig1} that at finite
temperatures there is no boundary (or phase transition) between the
states of systems located before or beyond QCP shown by the arrow.
Therefore, at elevated temperatures the properties of systems with
$x/x_{FC}<1$ or with $x/x_{FC}>1$ become indistinguishable, while
the particle-hole symmetry is restored \cite{pr}.
\begin{figure}[!ht]
\begin{center}
\includegraphics[width=0.45\textwidth]{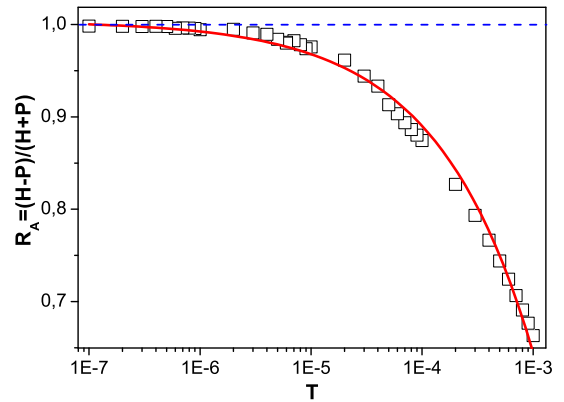}
\end{center}
\caption{$R_A$ as a function of the dimensionless temperature
measured in $E_F$. Our calculations are shown by squares. The solid
curve represents a fit $R_A\simeq a_0+a_1\sqrt{T}$.} \label{asymt}
\end{figure}
In fig. \ref{asymt} we present the asymmetry $R_A$ versus
dimensionless temperature. At low temperatures $T/E_F\leq 1$ the
asymmetry $R_A\simeq a_0+a_1\sqrt{T}$ and the symmetry is restored
at $T\simeq E_F$ since $R_A\simeq 0$. The function $R_A$ is of
universal form and the values of $a_0$ and $a_1$ are determined by
the location of the system at the quantum critical line shown in
fig. \ref{fig1}. We note that $a_0$ is given by the temperature
independent part $S_0$ of the entropy, $S(T\to0)\to S_0$
\cite{pr,vrshag,obz}.

One of the manifestations of the quasiparticle-hole symmetry on a
macroscopic scale is the symmetric electric conductivity. It is
straightforward to demonstrate that in Landau Fermi-Liquid the
differential conductivity $\sigma_{d}$ appearing in Ohm's law
\begin{equation}
\label{ohm}
dI\,=\,\sigma_{d} \,dV
\end{equation}
is a symmetric function of the voltage $V$, i.e.  $\sigma_{d}$ may
depend on the absolute value of $V$, but not on its sign. From eq.
\eqref{ohm} it follows that when the conductivity is a symmetric
function of voltage, reversing the sign of $V$ results in
$I\,\rightarrow\,-\,I$; electric current maintains its magnitude
and flows in the opposite direction. This reasonable feature may be
readily derived in the framework of the Landau Fermi Liquid theory.
Let us recall that the electric current can be expressed in terms
of the distribution function given by eq. \eqref{n} as \cite{deuts}
\begin{equation}
\label{I}
I(V)\,=\,\mathrm{const}\,\int\,d\varepsilon\,[n(\varepsilon-V)\,-\,n(\varepsilon)]
\end{equation}
Now, from eqs. \eqref{ohm} and \eqref{I} it immediately follows
that $\sigma_{d}$ is a symmetric function of voltage $V$. This
result is a direct consequence of the quasiparticle-hole symmetry
which is an inherent feature of the Landau Fermi-Liquid theory. The
symmetric conductivity has been observed so many times that it lead
to a perception that the conductivity can not be asymmetric.

This perception has been invalidated by recent experimental
observations where it was shown that at low temperatures the
electric conductivity in high-$T_{c}$ superconductors \cite{deuts}
as well as in some heavy fermion metals, such as $\rm CeCoIn_5$ and
$\rm YbCu_{5-x}Al_{x}$ \cite{exp1,exp2} is clearly asymmetric, this
asymmetry vanishing as the temperature increases. Evidently, the
asymmetry can not be explained in the framework of the Landau Fermi
Liquid theory since the quasiparticle-hole symmetry unavoidably
leads to the step function for the fermion distribution function in
the low temperature regime which, in turn results in a symmetric
conductivity. Therefore, in order to explain the asymmetric
conductivity, it is necessary to consider fermion systems more
general than Landau Fermi-Liquid.

Strongly correlated fermion systems may serve as one of the
examples of such systems. These systems have many novel phenomena
which have been observed experimentally \cite{ste,pr}; in our case
the most attractive is the low temperature asymmetric conductivity
\cite{deuts,exp1,exp2,pr}. This macroscopic effect finds its
explanation in the quasiparticle-hole asymmetry in theory of
strongly correlated fermion systems. Fundamental microscopic
interaction in this theory is fully symmetric with regard to
quasiparticles and holes but FCQPT at low temperatures causes the
spontaneous symmetry breakdown \cite{pr,vrshag}. Asymmetry is
caused by the simple fact that in strongly correlated fermion
systems, in contrast to Landau Fermi Liquid, the single-particle
energy $\varepsilon({\bf p})$ is $T$ dependent. Thus $n({\bf p},
T)$ given by eq. \eqref{n} does not reproduce the step function in
the low temperature limit \cite{ks,ksk}. The asymmetric part of the
conductivity,
\begin{equation}
\label{asymcond} \triangle
\sigma_{d}(V)\,\equiv\,{\sigma_{d}(V)\,-\,\sigma_{d}(-V)\over 2},
\end{equation}
can be calculated \cite{vrshag} and comparison with experiments
\cite{exp1,exp2} leads to good agreement. The estimate is
\begin{equation}
\label{ascond1} \triangle \sigma_{d}(V)\,\sim\,{V\over
2T}\,{p_{f}\,-\,p_{i}\over p_{F}}.
\end{equation}
From eq. \eqref{ascond1} it follows that a fairly large asymmetry
is obtained when $(p_{f}-p_{i})/p_{F}\approx 1$. When a magnetic
field $B$ is applied the Landau Fermi-Liquid behavior is restored,
particle-hole asymmetry is eliminated, and therefore the asymmetric
part of the differential conductivity disappears
\cite{pr,vrshag,obz}. In other words, the particle-hole symmetry is
macroscopically broken, or $CP$ is violated, in the absence of
applied magnetic fields. Conversely, the application of a magnetic
field restores both the particle-hole symmetry and the Landau Fermi
Liquid state. This agrees with the experimental facts collected in
measurements on $\rm YbCu_{5-x}Al_{x}$ \cite{pr,exp2}.

Now we are in a position to formulate our approach to the baryon
asymmetry of the universe. As it cools down, the universe behaves
as a non-Fermi Liquid; one of the manifestations of such systems is
a strongly correlated Fermi system exhibiting FCQPT at $T=0$ as is
shown in fig. \ref{fig1}. We suggest that this model describes the
particle-antiparticle content of the universe. As seen from fig.
\ref{asymt}, at finite temperatures baryon-antibaryon asymmetry
emerges as an inherent property of the system located above the
quantum critical line. The asymmetry results from the distortion of
the Fermi surface, in other words from the deviation of the
distribution function $n({\bf p})$ from the step function at low
temperature. At lowering temperature, the system approaches the
quantum critical line and, correspondingly, the asymmetry
increases. Details of this increase depend on the model chosen; the
very existence and the universal qualitative behavior of the
asymmetry is of great significance.

The picture for explaining baryon asymmetry emerging from the above
is as follows. The initial excited state corresponding to the big
bang with extremely high temperature possesses matter-antimatter
symmetry. At the end of the inflation stage the dark matter emerges
producing baryons (holes) and anti-baryons (quasiparticles). At
this stage the temperature is high and the chemical potentials of
baryons and antibaryons are zero, so that the asymmetry is also
zero. As the temperature drops the state with the baryon (hole)
asymmetry is formed. As a result, the universe approaches the
quantum critical line which corresponds to the state with the
maximum baryon (hole) asymmetry as it is seen from fig.
\ref{asymt}. This state is the eigenstate of the fully symmetric
microscopic Hamiltonian with the eigenvalue lower than the one for
the state with matter-antimatter symmetry and contains the visible
matter - baryons (holes), and the dark matter in the vicinity of
the visible matter. In our approach the visible matter is
represented by the excitations (holes) of the dark matter. As there
is almost no interaction between holes and the Fermi sea, we
conclude that there is no direct interaction between dark matter
and visible matter, that is the interaction if it exists is very
weak. Then, proceeding along the universe-non-Fermi Liquid analogy,
since it takes about ten particles to create one quasiparticle
\cite{landau}, we estimate the ratio of the mass of visible matter
to the mass of dark matter to be of order of ten which is close to
the observed value $\Omega_{DM}/\Omega_{b}\approx 5$ \cite{BBB,
pdg}. The ground state of the universe which we identify with the
dark energy is interpreted in our approach as the vacuum.

Another result which comes as a bonus of our universe-non-Fermi
Liquid analogy is the high entropy of the universe. As it is seen
from fig. \ref{figfc}, $n({\bf k},T)$ of holes (the visible matter)
even at $T\to0$ is non-integer, $0<n({\bf k},T)<1$. The entropy
$S(n({\bf p},T))$  is given by the well-known expression
\cite{landau}
\begin{eqnarray}
S(n({\bf p},T))&=& -2\int[n({\bf p},T) \ln (n({\bf p},T))+(1-n({\bf p},T))
\nonumber \\
&\times&\ln (1-n({\bf p},T))]\frac{d{\bf p}}{(2\pi) ^3}.\label{FL3}
\end{eqnarray}
It follows from fig. \ref{figfc} and eq. \eqref{FL3} that the
entropy of the system is finite at $T\to0$: $S(T\to0)\to S_0$.

Let us introduce $S_B$, entropy per baryon, as  $S/x$ where $S$ is
given by eq. \eqref{FL3} and $x$ is the number density of baryons.
Then from the eq. \eqref{FL3} it follows that $S_B\sim1$. This
observation immediately explains the high entropy of the visible
matter \cite{entr}. We also conjecture that the observed violation
of $CP$-symmetry, leading to the violation of $T$-symmetry and
making the finite term of the entropy, $S(T\to0)=S_0$, may resolve
the well-known problem of the time arrow.

It might seem that the presence of the Fermi level contradicts the
relativistic invariance and the $CPT$ theorem since the Fermi level
and the Fermi sphere related to it are not transformed according to
the invariance. As we shall see, at high energies the degrees of
freedom related to the Fermi sphere become irrelevant and the
relativistic symmetry is preserved in our approach. Let us assume
that there are substantial gaps in the energy scales separating
different states of the evolution of the Universe. One of these
gaps separates the state behind FCQPT which produces $CP$ violation
and the symmetrical state laying above FCQPT. Consider the linear
response function $\chi({\bf q},\omega)$ (see, e.g., \cite{PinNoz})
\begin{equation}\label{chi}
\chi({\bf q},\omega)=\sum_n
|\rho(q)_{n0}|^2\left[\frac{2\omega_{n0}}
{(\omega+i\gamma)^2-\omega_{n0}^2}\right].
\end{equation}
Here $\rho({\bf q})$ describes the fluctuations of the
quasiparticle density, $\omega_{n0}=E_n-E_0$ is the difference
between the ground state and the excited state energies. In our
case $E_n$ are the energy of excited states behind FCQPT.  If
$\omega$ is sufficiently high so that $\omega\gg (E_n-E_0)$ then
eq. \eqref{chi} defines the linear response function of
noninteracting particles \cite{PinNoz}
\begin{equation}\label{chi1}
\chi({\bf q},\omega\to\infty)=\frac{x q^2}{m\omega^2},
\end{equation}
where $m$ is the mass of quasiparticle at 
$\omega\gg \omega_{n0}$. In our case, it means that at these high
energies the system is located in the region of its phase diagram
located well above the critical line shown in fig. \ref{fig1} where
the $CP$ symmetry is restored. Relativistic invariance is not yet
restored since the very existence of quasiparticles ensures that
Fermi sphere still remains relevant surface in phase space. Now we
take into account that the linear response function of this system
is again given by eq. \eqref{chi} with new $\omega^{\prime}_{n0}$.
Again at $\omega\gg \omega^{\prime}_{n0}$ the response function is
given by eq. \eqref{chi1} formed by particles the existence of that
is not related to any Fermi sphere. Going along this way, we ascend
a level at which the inflation takes place and the relevant degrees
of freedom are now not quasiparticles but particles of Standard
Model.  As a result, we conclude that at elevated energies $\omega$
the irrelevant degrees of freedom vanish and both the relativistic
invariance and the $CPT$ symmetry emerge. The details of
reappearing of relativistic invariance are defined by the concrete
model chosen to describe baryogenesis.

The attractive feature of the scenario discussed above is in its
``conservative'' character - in order to introduce the
matter-antimatter asymmetry it may suffice to suggest a mechanism
based on analysis of quantum phase transitions in Fermi systems.
There is no need to introduce baryon number non conservation or
$CP$ violating interactions, as well as to invoke any extension of
the standard cosmological model - ordinary quantum mechanics and
quantum statistics applied to a multi fermion system guarantee that
the system starts from the symmetric state and at decreasing
temperatures arrives at a maximally asymmetric state. This
universal feature is present in strongly correlated Fermi systems
with the fermion condensation quantum phase transitions; the
details depend on the concrete microscopic Hamiltonian postulated.
We note that typical current-current interactions lead to the
formation of flat bands \cite{rice}. These located at the Fermi
surface, being topologically protected from interaction and other
disturbances, lead to the robustness of the generic properties of
the quantum vacuum generated by their existence
\cite{vol,vol1,vol2}. Because of the universal features of our
model we have not concentrated on a particular picture that follows
from some concrete dynamics, nor pursued the best quantitative
description as the goal by itself. Rather, we demonstrate an
opportunity to explain baryon asymmetry from the very general
physics principles. Let us stress that the quasiparticle-hole
asymmetry which manifests itself at the macroscopic scale is
observed in experiments on heavy fermion metals and analyzed
theoretically is one of the few analogies of particle-antiparticle
asymmetry observed in the universe, and thus deserves attention.

We grateful to A. D. Kaminker, V. A. Khodel, D. A. Varshalovich and
D. G. Yakovlev for valuable discussions. This work is supported by
U.S. DOE, Division of Chemical Sciences, Office of Basic Energy
Sciences, Office of Energy Research, AFOSR and the RFBR \#
09-02-00056.

\end{document}